\documentclass[%
 reprint,
amsmath,
amssymb,
aps,
pra,
]{revtex4-2}

\usepackage{graphicx}
\usepackage{dcolumn}
\usepackage{bm}
\usepackage{hyperref}

\usepackage{subcaption}
\usepackage{caption}

\begin{document}


\title{Simultaneous SI-traceable measurement of the amplitude and the ellipticity of microwave fields with cold Rydberg atoms}


\author{Anthony El Bekai}
\author{Romain Granier}
\author{Miguel Angel Cifuentes Marin}
\author{Cédric Blanchard}
\author{Nassim Zahzam}
\author{Yannick Bidel}
\author{Alexandre Bresson}
\author{Alexis Bonnin}
\author{Sylvain Schwartz}
\affiliation{DPHY, ONERA, Université Paris-Saclay, 91120 Palaiseau, France}


\date{\today}

\begin{abstract}
We present a method for the simultaneous measurement of the amplitude and the ellipticity of microwave fields using trap-loss spectroscopy based on Rydberg excitation of a cold ensemble of $^{87}$Rb atoms. The measurement consists of probing the energy spectrum of the $61S_{1/2}$ to $61P_{1/2}$ transition in the presence of microwave, which depends on the electric field amplitude $|E|$ and the third Stokes parameter $|S_3|$. We use a millimeter-size magneto-optical trap to perform strongly-localized measurements, which mitigates the effect of microwave field inhomogeneities. This allows us to work with large Rabi frequencies and thus resolve small deviations from linear or circular states of polarization. Taking into account the effect of non-resonant couplings to neighboring Rydberg states, we measure $|E|$ and $|S_3|$ with an accuracy on the order of 1 percent in the Autler-Townes regime where the measurements can be traced to the international system of units. By studying the visibility of the spectral features as a function of the polarization of the excitation laser beams, we extract some information on the angle of arrival of the microwave field onto the atoms, with good agreement between theory and experiment.
\end{abstract}


\maketitle


\section{Introduction}

Rydberg atoms are promising candidates for electromagnetic field sensing owing to their unique properties such as large electric dipole moments, long lifetimes and transition frequencies spanning a large spectrum \cite{sedlacek2012microwave,holloway2014broadband}. In particular, they have the potential to perform SI-traceable measurements of the amplitude of the electric field interacting with the atoms \cite{holloway2014broadband,Schlossberger2024} in a fully dielectric sensor head \cite{simons2018fiber}. Typically, in the case of a linearly-polarized microwave (MW) field addressing an $S_{1/2}$ to $P_{1/2}$ or $P_{3/2}$ transition, it can be shown \cite{cloutman2024polarization,yuan2024isotropic} that the measured Autler-Townes splitting $|\langle \mathbf{\hat{d}}\cdot \mathbf{E}\rangle|$ is independent of the orientation of the polarization, allowing an absolute measurement of $|\mathbf{E}|$ based on the tabulated value of the dipole matrix element. The situation becomes more complex for different atomic transitions and/or different states of polarization of the electric field. For example, in reference \cite{sedlacek2013atom}, a $P_{3/2}$ to $D_{5/2}$ transition was used to measure the orientation of a linearly-polarized MW field. For this, the amplitude of the two lines of an Autler-Townes doublet was compared with the amplitude of a central spectral line resulting from the absence of $\pi$-coupling between the extremal sub-levels of $D_{5/2}$ and $P_{3/2}$. Along a similar line, the authors of \cite{Jiao2017} infer the orientation of a 100~MHz linearly-polarized radio-frequency (RF) field by comparing the amplitudes of different electromagnetically-induced transparency (EIT) peaks involving Rydberg states dressed by the non-resonant RF field. An alternative method relies on a Rydberg-atom-based mixer \cite{wang2023precise}, where the orientation of a linear polarization is measured by looking at the amplitude of the beat note with a local linearly-polarized oscillator, although the need for a local oscillator makes the system less SI-traceable and dielectric. This has been generalized to measure any arbitrary state of polarization using heterodyning with 3 independent orthogonal local oscillators \cite{elgee2024complete}.

In references \cite{Chopinaud2021,Chopinaud2024}, it was shown that in the presence of an elliptically-polarized field addressing an $S_{1/2}$ to $P_{1/2}$ transition, four peaks could be observed owing to different coupling strengths between different $m_J$ states, which could be used to perform an SI-traceable measurement of the ellipticity of the incident polarization based on frequency measurements only. This opened the way to a series of recent articles harnessing Zeeman sublevels of Rydberg states to infer or control the state of polarization of an incident field. In reference \cite{Kurdak2025}, a bias magnetic field in a cold and spin-polarized atomic sample was used to lift the degeneracy between $m_J$ sublevels and selectively address in frequency $\pi$, $\sigma_+$ and $\sigma_-$ transitions, allowing to produce a pure state of polarization for the incident MW field using in-vacuum electrodes. In reference \cite{wang2026self}, EIT in a spin-polarized cold atom ensemble was used to probe the eigenstates resulting from the coupling between Zeeman sublevels of $61S_{1/2}$ and $61P_{3/2}$, and a multiparameter non-linear fit allowed to infer the polarization of the incident MW field. In reference \cite{Holloway2026}, the authors used a long-wavelength $D_{5/2}$ to $P_{3/2}$ Rydberg transition in a vapor cell to limit MW field inhomogeneities and better resolve the EIT peaks at large MW power, while measuring the ellipticity of the MW field from the eigenfrequencies resulting from the coupling between different $m_J$ states. Finally, reference \cite{Kjaergaard2026} gives a theoretical model and an experimental proof-of-principle of an ellipticity measurement based on the eigenfrequencies of coupled $m_J$ states for various types of transitions ($J=1/2$ to $J'=1/2$, but also 1/2 to 3/2, 3/2 to 3/2 and 3/2 to 5/2) in a vapor cell.

In the above approaches, the quality of the ellipticity measurement is set by the ability to experimentally resolve the different eigenstates, which depends on the width of the spectral features and on the difference between the eigenfrequencies. For a given MW power, the latter becomes increasingly small as the polarization approaches a linear or a circular state, which makes the estimation of the residual ellipticity increasingly difficult, as observed in \cite{Kjaergaard2026}. One way to overcome this problem is to increase the MW power, but the spectral features tend to blur at high power because of spatial inhomogeneities of the electric field \cite{Chopinaud2021}, unless a long-wavelength transition is used \citep{Holloway2026}.

In this paper, we build on the previous work described above to quantitatively assess the performance of the SI-traceable measurement of both the amplitude and the ellipticity of a MW field using an unpolarized ensemble of cold atoms. For this, we use the simple $S_{1/2}$ to $P_{1/2}$ transition between neighboring Rydberg states, which allows us to perform metrological measurements thanks to a detailed and precise description of the physical system. In a first approach, the system’s energy spectrum is simply modeled as two independent 2-level systems coupled through the two $\sigma_+$ and $\sigma_-$ components of the MW field from which we extract $|E|$ and $|S_3|$ using a single and unambiguous value of the electric dipole. We implement this method using trap-loss spectroscopy \cite{Duverger2024Metrology} in a cold atom cloud trapped in a volume of a few mm$^3$. Interestingly, this approach does not require a long-wavelength transition and is compatible with most radar frequency bands. Owing to this strongly localized measurement, we obtain well-resolved spectral features even at strong MW power. This allows to detect and measure small deviations from linearly or circularly polarized fields where eigenenergies are close to degeneracy. As the microwave power increases, it becomes necessary to take into account the influence of neighboring non-resonant Rydberg states to maintain an accurate estimate of $|S_3|$, which we do perturbatively. This allows us to determine the amplitude and the ellipticity at the 1\% level over a 15 dB dynamic range in the Autler-Townes regime where the measurement can be traced to the international system of units \cite{holloway2014broadband}. We also study the influence of the polarization of the Rydberg excitation lasers on the visibility of the spectral features, with very good quantitative agreement between theory and experiment, from which we infer four possible angles of arrival of the microwave onto the atoms, due to the symmetries of the system.

\section{Polarization dependent spectrum of a resonantly coupled $|nS_{1/2}\rangle\leftrightarrow|n'P_{1/2}\rangle$ Rydberg transition}\label{theory}

In this paper, we focus on the Rydberg transition between states $|nS_{1/2}\rangle$ and $|n'P_{1/2}\rangle$. This constitutes a 4-level atomic system whose basis of magnetic sub-levels is given by $\big\{ |s, m_J=\pm\frac{1}{2}\rangle;|p, m_J=\pm\frac{1}{2} \rangle \big\}$. The states $|nS_{1/2}\rangle$ and $|n'P_{1/2}\rangle$ are resonantly coupled by an oscillating electric field $\vec{E}(t)= |E| \cos (\omega t) \vec{\epsilon}$ with an arbitrary polarization state described by the complex unitary vector $\vec{\epsilon}$. We choose for the quantization axis $\vec{e}_z$, the propagation axis of the microwave, and we express the polarization state $\vec{\epsilon}$ in the spherical basis $\big\{ \vec{\epsilon}_{0} = \vec{e}_{z};\vec{\epsilon}_{+} = -(\vec{e}_{x}+i\vec{e}_{y})/\sqrt{2};\vec{\epsilon}_{-} = (\vec{e}_{x}-i\vec{e}_{y})/\sqrt{2} \big\}$. In that basis, we then express the electric field components as $\big\{ E_0=E_z=0 ; E_+ =(E_x+iE_y)/\sqrt{2} ; E_- =-(E_x-iE_y)/\sqrt{2} \big\}$, with $E_x = |E| \cos \left(\frac{\theta}{2}\right)$, $E_y = |E| \sin \left(\frac{\theta}{2}\right) e^{-i\phi}$ and where $\theta \in [0,\pi]$ and $\phi \in [0,2 \pi[$.

\begin{figure}
    \includegraphics[width=1\linewidth]{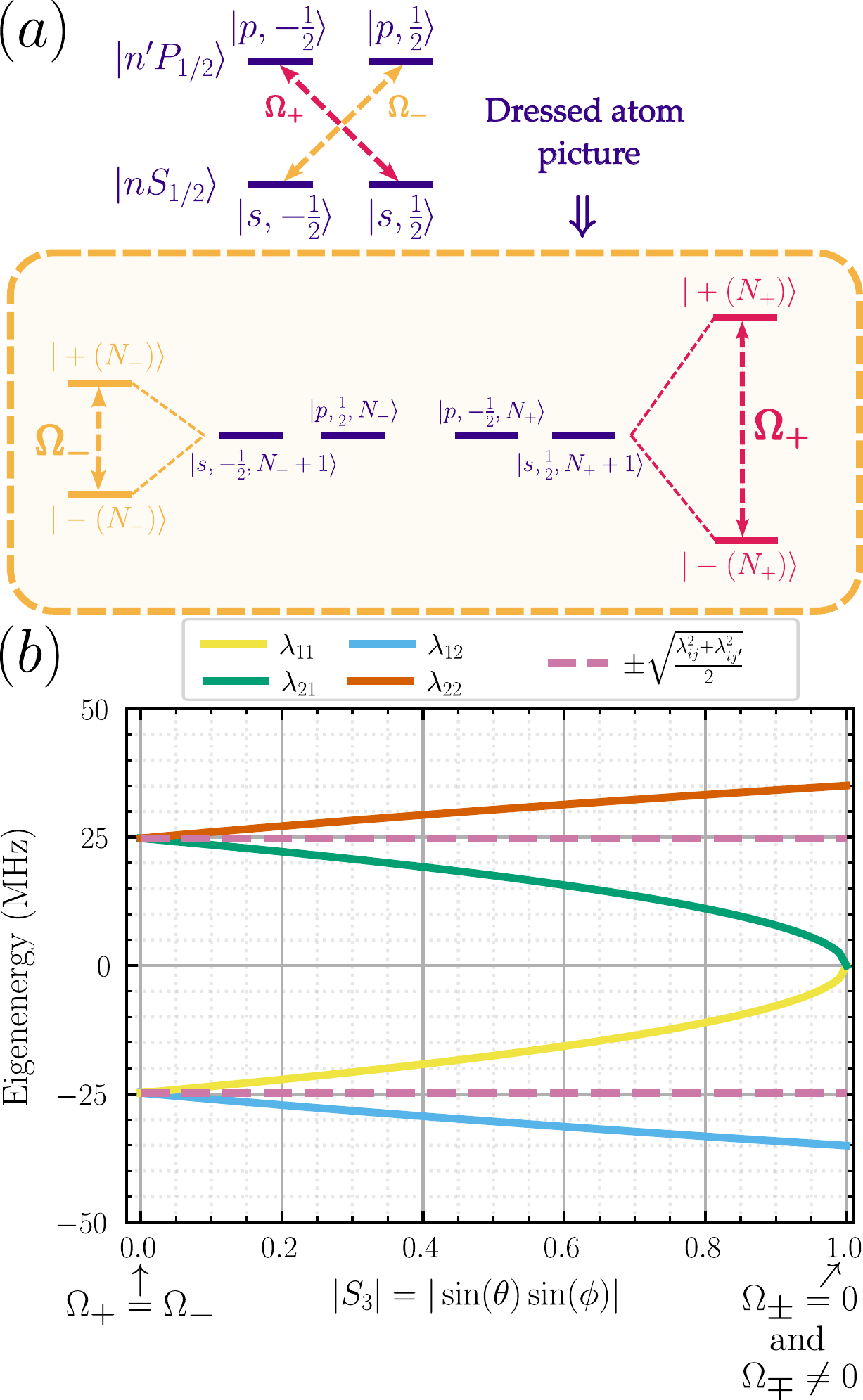}
    \caption{(a) The interaction between the four level atomic system and the microwave field results in two $\sigma_{+,-}$ transitions of strength $\Omega_{+,-}$. In the dressed atom picture, this leads to two Autler-Townes doublets whose splittings depend on the microwave polarization through the relative amplitudes between $\Omega_+$ and $\Omega_-$. For the bare states, $N_{+,-}$ denotes the number of microwave photons.  (b) Eigenenergies (solid colored lines), in units of MHz, as a function of the absolute value of the  third Stokes parameter $|S_3|$ obtained from Eq.(\ref{eq:eigenvaluesvsS3}) for a constant electric field amplitude $|E|=3$ V/m with $d = 1291.5 ~ e a_0$ considering the n=61 Rydberg states. The two root-mean-square values $\sqrt{(\lambda_{21}^2 + \lambda_{22}^2)/2}$ and $- \sqrt{(\lambda_{11}^2 + \lambda_{12}^2)/2}$} (dashed pink lines) are plotted to illustrate a constant `average' Autler-Townes splitting $\omega_{AT}$ (\textit{cf.} Eq.(\ref{eq:splitting})) and thus a constant field amplitude $|E|$.  
    \label{fig:eigenvalues}
\end{figure}

In the dipole and rotating wave approximations, considering resonant MW coupling and neglecting residual magnetic fields and AC Stark shifts from neighboring Rydberg states, we can express the effective total Hamiltonian $\hat{H}$ as the following block diagonal matrix in the native J-basis:
\[
\hat{H} = \frac{\hbar}{2}
\renewcommand{\arraystretch}{1.6}
\setlength{\arraycolsep}{6pt}
\raisebox{2ex}
{$
\begin{array}{c@{\hspace{-10pt}}c}
&
\begin{array}{cccc}
{|s,\frac12}\rangle &
{|p,-\frac12}\rangle &
{|s,-\frac12}\rangle &
{|p,\frac12}\rangle
\end{array}
\\[1ex]
\begin{array}{c}
\langle{s,\frac12}| \\
\langle{p,-\frac12}| \\
\langle{s,-\frac12}| \\
\langle{p,\frac12}|
\end{array}
&
\left[
\begin{array}{cccc}
0 & \Omega_+ & 0 & 0 \\
\Omega_+^* & 0 & 0 & 0 \\
0 & 0 & 0 & -\Omega_- \\
0 & 0 & -\Omega_-^* & 0
\end{array}
\right]
\end{array}
$}
\]
where the Rabi coupling strengths for $\big\{ \pi ; \sigma_+ ; \sigma_- \big\}$ transitions are respectively given by $\big\{ \hbar\Omega_0 = -dE_0 ; \hbar \Omega_+ = -\sqrt{2}dE_+ ; \hbar \Omega_- = -\sqrt{2}dE_- \big\}$, where $d = -e\mathcal{R}/3$ with $e$ the elementary charge and $\mathcal{R} = \int_{0}^{\infty} R_{n,l=0}^*(r) R_{n',l=1}(r) r^3 dr$ the radial matrix element. The associated spectrum contains four eigenenergies whose expressions are directly given by $\hbar/2\times\left\{ \pm|\Omega_+| ; \pm|\Omega_-| \right\}$. In the dressed atom picture (see Fig.\ref{fig:eigenvalues}(a)), the polarization dependence of the system's energy spectrum can be easily interpreted as follow: the right and left circular components of the microwave field create two independent Autler-Townes doublets, whose splitting is given by $\Omega_+$ and $\Omega_-$ respectively. In the general case of an elliptical polarization, \textit{i.e.} $\Omega_+ \neq \Omega_-$, the four eigenvalues are non-degenerate. For a linear polarization, \textit{i.e.} $\Omega_+ = \Omega_-$, the two Autler-Towes splittings are equal and we thus obtain two eigenvalues, each of multiplicity two. For a circular polarization, \textit{i.e.} $\Omega_+ = 0$ or $\Omega_- = 0$, the spectrum consists of three eigenvalues, a central twofold degenerate eigenvalue and the non-degenerate AT doublet.

From this simple model, we deduce that the spectrum contains information on both the amplitude and the polarization state of the electric field. More specifically, we use the Stokes parameters to parametrize the polarization within the Poincaré sphere formalism (with polar angles $\theta$ and $\phi$) where the ellipticity angle $\chi$ is linked to the third Stokes parameter $S_3$ by the relation $S_3= \sin (2\chi) = \sin \theta \sin \phi $. The four eigenvalues can then be rewritten in the following form:        
\begin{equation}
    \lambda_{ij} = (-1)^i \frac{dE}{2\hbar} \sqrt{1 + (-1)^j |S_3| }
\label{eq:eigenvaluesvsS3}
\end{equation}
with $\left(i;j\right) \in (1;2)$. Without any previous knowledge of the polarization of the electric field, one can derive analytical relationships from the eigenvalues expressions for $|E|$:

\begin{equation}\label{eq:splitting}
\omega_{AT} = \frac{d \, |E|}{\hbar} = \sqrt{\frac{\lambda_{21}^2 + \lambda_{22}^2}{2}} + \sqrt{\frac{\lambda_{11}^2 + \lambda_{12}^2}{2}}
\end{equation}

or equivalently:

\begin{equation}\label{eq:splitting_v2}
\omega_{AT} =
\sqrt{\frac{|\Omega_+|^2}{2} + \frac{|\Omega_-|^2}{2}} = \sqrt{\frac{(\lambda_{22}-\lambda_{12})^2}{2} + \frac{(\lambda_{21}-\lambda_{11})^2}{2}}
\end{equation}

and for $|S_3|$:
\begin{equation}\label{eq:S3}
\begin{split}
|S_3| &= |\sin (2\chi)| \\
&= \frac{2\hbar^2}{d^2 |E|^2} \left[ (\lambda_{22}-\lambda_{12})^2 - (\lambda_{21}-\lambda_{11})^2 \right]\\
&= \frac{ \left| | \Omega_+|^2 - |\Omega_-|^2 \right| }{|\Omega_+|^2 + |\Omega_-|^2}
\end{split}
\end{equation}
Note that $\omega_{AT}$ in Eq.(\ref{eq:splitting}) corresponds to the mean value of the two root-mean-square values $\sqrt{(\lambda_{21}^2 + \lambda_{22}^2)/2}$ and $- \sqrt{(\lambda_{11}^2 + \lambda_{12}^2)/2}$ as $\lambda_{12} \leq \lambda_{11} \leq 0$. This quantity coincides with the usual Autler-Townes splitting in the case of a linear polarization.  

To get more insight about the dependence of the spectrum to the polarization ellipticity, we represent in Fig. \ref{fig:eigenvalues}(b) the four eigenenergies as a function of $|S_3|$, obtained from Eq.(\ref{eq:eigenvaluesvsS3}), considering a constant electric field amplitude $|E|$. From linear ($|S_3|=0$; 2 eigenenergies), through elliptic ($ 0<|S_3|<1 $; 4 eigenenergies)), to circular ($|S_3|=1$; 3 eigenenergies) polarization, the `average' Autler-Townes splitting $\omega_{AT}$ remains constant (see dashed lines). In contrast, the difference between $|\Omega_{\pm}|^2=(\lambda_{22}-\lambda_{12})^2$, corresponding to the two external eigenvalues, and $|\Omega_{\mp}|^2=(\lambda_{21}-\lambda_{11})^2$, corresponding to the two internal eigenvalues, and thus the ellipticity $\chi$, increases.

This simple model suggests that a measurement of the RF-coupled system's energy spectrum through spectroscopic technique allows the simultaneous estimation of the RF field amplitude $|E|$ and ellipticity $|S_3|$ using Eq.(\ref{eq:splitting_v2}) and Eq.(\ref{eq:S3}), in a way which is traceable to the international system of units.

\section{Experimental measurement of $|E|$ and $|S_3|$}\label{Experiment}

To evaluate this method, we study and analyze the spectra obtained by resonantly coupling the $|r_1\rangle=|61S_{1/2}\rangle\leftrightarrow|r_2\rangle=|61P_{1/2}\rangle$ transition with a microwave field at 15.973 GHz for three different polarization states.

\begin{figure}
	\includegraphics[width=0.9\linewidth]{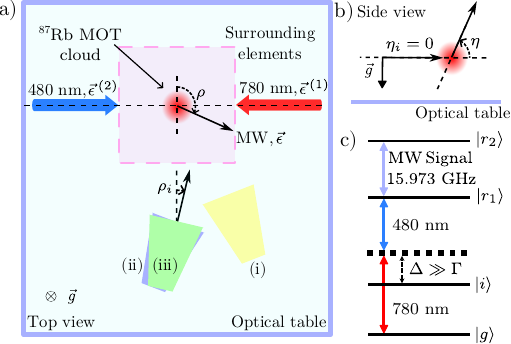}
	\caption{Schematic view of the experimental setup: a) Top view where (i), (ii), (iii) illustrate the positions and orientations of the horn antenna for quasi-linear, quasi-circular and elliptical configuration respectively; 	
	$\vec{\epsilon}$, $\vec{\epsilon}^{\,(1)}$ and $\vec{\epsilon}^{\,(2)}$ correspond to the polarization state of the MW signal, the 780 nm laser beam and the 480 nm laser beam respectively. Initially, the MW polarization state is aligned with the vertical $\vec{g}$; b) Side view, we represent the in-plane angle $\rho$, the  out-of-plane angle $\eta$, at the atoms location and their initial directions at the horn output (indexed by i). (see Section \ref{rydberg_polarization}); c) Energy level diagram and couplings.}
	\label{fig:Exp_setup}
\end{figure}

The experimental setup and the level diagram are depicted in Fig. \ref{fig:Exp_setup}. Our experiment is based on a Magneto-Optical Trap (MOT) of $\sim 5 \times 10^7$ atoms of $^{87}\text{Rb}$ at a temperature of $T \sim 500 ~ {\mu K}$ (see more details in \cite{Duverger2024Metrology}). We apply a magnetic field gradient of 22 G/cm that confines the atoms in a volume of $1.8\times 1.6 \times 1.6$~mm$^3$. We consider the trapped atomic sample in the MOT to be unpolarized on average. The Rydberg excitation is performed with an effective two-photon transition between the ground state $|g \rangle = |5S_{1/2}, F=2\rangle$ and the Rydberg state  $|61S_{1/2}\rangle$ via the intermediate states $|i \rangle = |5P_{3/2},F'\rangle$, $F' \in \{1,2,3\}$, with a blue detuning of $\Delta/2\pi = 3.2$ GHz above the $|5P_{3/2}, F'=3\rangle$ state. The lasers at 780 nm and 480 nm are counterpropagating with controllable polarization states $\vec{\epsilon}^{\,(1)}$ and $\vec{\epsilon}^{\,(2)}$. We phase-modulate a portion of these lasers to lock a sideband using the Pound-Drever-Hall technique on an ultrastable Fabry-Perot cavity. This allows us to reach linewidths in the 100 Hz regime and a tunability of a few hundreds of MHz. To record the spectra keeping $\Delta$ constant, we scan the frequency of the 480~nm laser at a rate of 2~MHz/s which minimizes dynamical broadening because the response time of the MOT of $\sim$ 1~s. To generate the microwave electric field, we use a horn antenna that emits a vertical linearly polarized field at its output. In the presence of the surrounding environment constitued by numerous reflecting and diffracting elements for the microwave signal at the wavelength $\lambda\sim1.9~$cm, we empirically find three horn positions for which the effective polarization measured by the atoms is respectively (i) quasi-linear, (ii) quasi-circular and (iii) elliptical as illustrated in Fig. \ref{fig:Exp_setup}.      

\begin{figure}
	\includegraphics[width=0.9\linewidth]{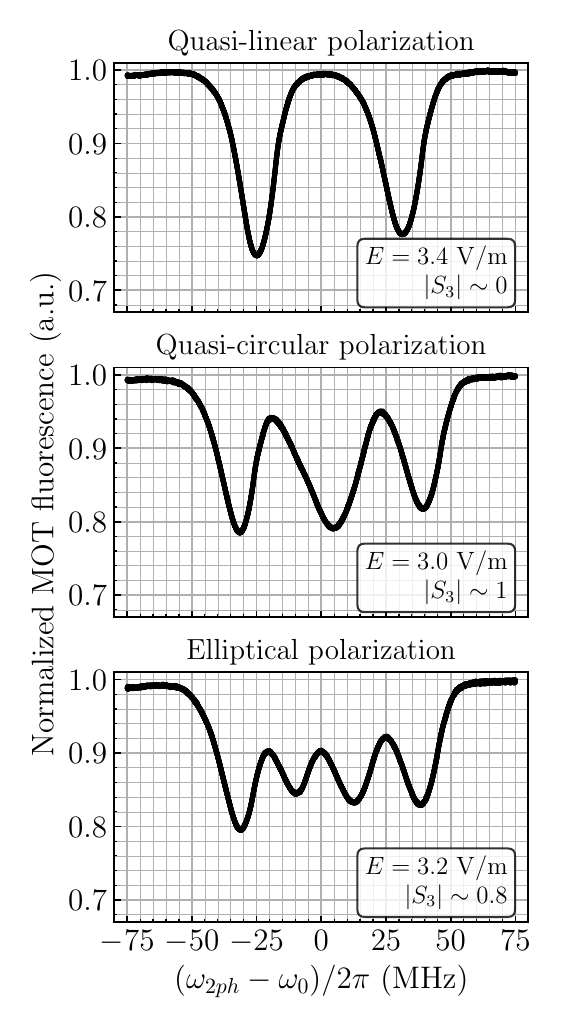}
	\caption{Trap-loss spectra for three distinct microwave polarizations as a function of the two-photon laser detuning. Linear, circular and elliptical microwave polarizations result in 2, 3 and 4 resonances respectively. We extract $|E|$ using Eq.\eqref{eq:splitting} and give a first estimate of $|S_3|$ directly using Eq.\eqref{eq:S3}.}
	\label{fig:3spectra}
\end{figure}

The measurement protocol we use is based on trap-loss spectroscopy, as explained in \cite{Duverger2024Metrology}. Experimentally, we observe a visible drop in the MOT fluorescence signal as the effective loss rate induced by the blackbody ionization from the Rydberg states is similar to the inner loss rate of the trap. On Fig.\ref{fig:3spectra}, we show three normalized trap-loss spectra obtained in the three spatial configurations mentioned above. As expected from the simple model of section \ref{theory}, we clearly observe two, three and four resonances for linear, circular and elliptical polarization respectively.

The quality of the measurement of $|S_3|$ is related to the capability to resolve and precisely determine the center frequencies of the spectral features. At low field amplitude, the frequency resolution is limited by the linewidth of the spectral features, on the order of 15~MHz (FWHM), resulting from the effects of cooling light and the magnetic field gradient \cite{Duverger2024Metrology}. Increasing the field amplitude allows to increase the splittings and thus to lower the relative uncertainty on their estimation. This is made possible in our case by the fact that our mm-sized atomic cloud performs a measurement of the MW field which is well-localized on the scale of the MW wavelength, limiting the broadening effect from the spatial inhomogeneities of the MW field. This allows in some situations resolving quasi-degenerate eigenstates and thus detecting possible deviations from perfectly linearly or circularly polarized microwaves. Importantly, increasing the electric field amplitudes also increases the energy shifts induced by off-resonant couplings from neighboring Rydberg states refered to as AC Stark shifts. These shifts must be accounted for to obtain an accurate estimate of $|S_3|$, especially when the microwave splittings become non-negligible with respect to the detunings from neighboring Rydberg transitions: in our case, the closest off-resonant transition, namely $|61S_{1/2}\rangle\leftrightarrow|61P_{3/2}\rangle$, is blue detuned by 437 MHz from the targeted $|61S_{1/2}\rangle\leftrightarrow|61P_{1/2}\rangle$ transition. In the following, we evaluate the energy shifts for each of the four $\big\{ |s, m_J=\pm\frac{1}{2}\rangle;|p, m_J=\pm\frac{1}{2} \big\}$ states by summing the contributions of all the allowed $\{\sigma_+,\sigma_-\}$ transitions of amplitude $\hbar\Omega_{\pm,i}^2/(4\Delta_i)$ between the atomic manifolds of $|61S_{1/2}\rangle$ and $\{|61P_{3/2}\rangle;|60P_{1/2}\rangle;|60P_{3/2}\rangle\}$ and between $|61P_{1/2}\rangle$ and $\{|62S_{1/2}\rangle;|60D_{3/2}\rangle;|59D_{3/2}\rangle\}$.  By doing so, we add to $\hat{H}$ $|S_3|$-dependent diagonal terms with an $|E|^2$ scaling and obtain the refined Hamiltonian $\hat{H}'$ given in appendix \ref{Appendix:B}.
\begin{figure}
	\includegraphics[width=0.9\linewidth]{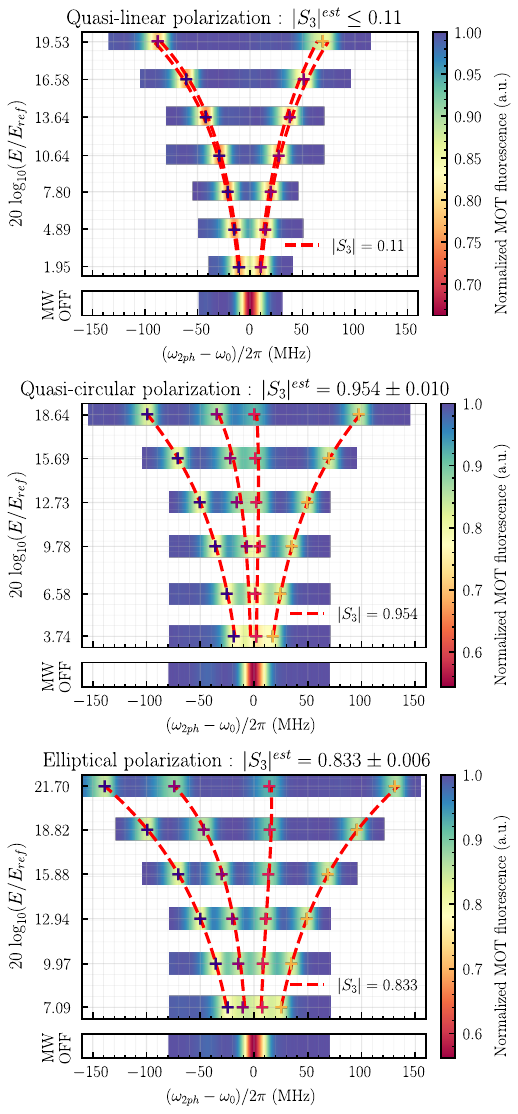}
	\caption{Trap-loss spectra colormaps for the three distinct microwave polarizations and for different values of the field amplitude $|E|$. The vertical axes correspond to $|E|$, estimated at the atomic cloud position directly from Eq.\eqref{eq:splitting_v2}, in power log scale relative to $E_{ref} = 1$ V/m. The bottom subsets show the trap-loss profile without microwave field used as frequency reference. The cross symbols represent the center frequency values extracted from Gaussian fits of the experimental data. We estimate $|S_3|$ as described in the main text and red dashed lines show calculated eigenvalues for these $|S_3|^{\mathrm{est}}$ values. For each polarization, the best estimation of the third Stokes parameter $|S_3|^{\mathrm{est}}$ is given with an associated uncertainty corresponding to a 95\% confidence interval, except for the quasi-linear case where the upper bound is set by the width of the spectral features (see text for more details).}
	\label{fig:ColorMaps_S3est}
\end{figure}

For the three states of polarization considered, we perform trap-loss measurements for several electric field amplitudes typically ranging from 1.25 to 12.2 V/m as shown in Fig.\ref{fig:ColorMaps_S3est}. We average each spectrum over 3 pairs of scans of opposite directions to mitigate the influence of the MOT dynamics \cite{Duverger2024Metrology}. We extract the center frequencies of each resolved resonance by multi-peak gaussian fits. We then estimate the microwave field characteristics as follow: first, we derive the electric field amplitude $|E|$ at the atomic cloud position for each applied power using Eq.\eqref{eq:splitting_v2} and given $d = 1291.5 ~ ea_0$ (where $e$ is the elementary charge and $a_0$ the Bohr radius). For this estimation, we neglect the AC Stark shifts as they induce relative errors on the estimation of $|E|$ smaller than 1 \% (see appendix \ref{Appendix:B}). We then numerically diagonalize the Hamiltonian $\hat{H'}$ based on the estimated field amplitudes. Finally, we minimize the distance between these eigenenergies and the corresponding experimental values with $|S_3|$ as the only free parameter (see appendix \ref{DataAnalysis} for more details).   

\begin{figure}
	\includegraphics[width=0.95\linewidth]{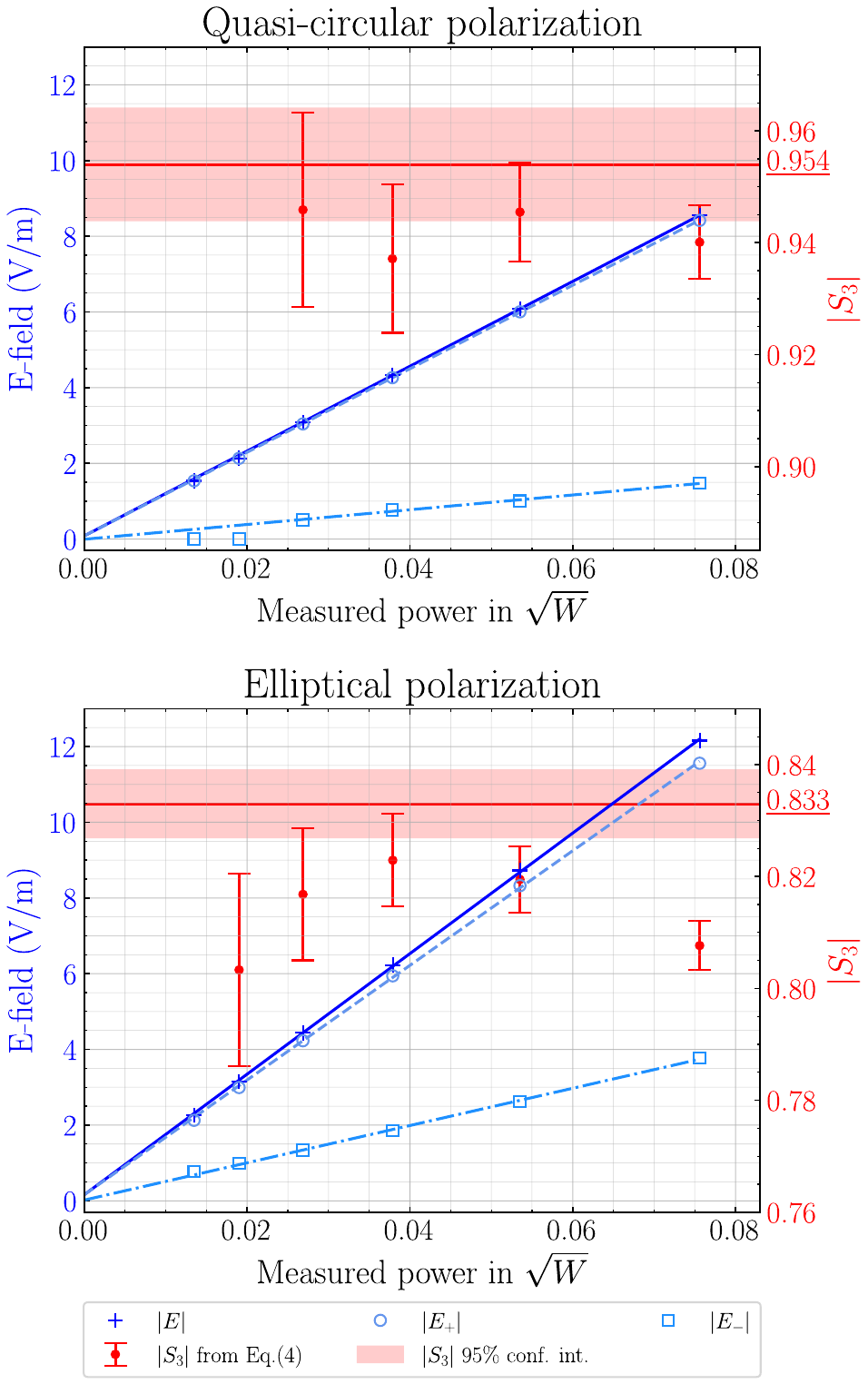}
	\caption{Estimation of the field amplitude and ellipticity as a function of the square root of the RF power sent at the input of the horn antenna for two polarization states. The blue crosses, respectively circles and squares, correspond to the total field amplitude, respectively to the amplitudes of the spherical components in units of V/m (left axis). Solid, dashed and dot-dashed blue lines are linear fits to the data of these 3 quantities. The red dots correspond to the estimations of $|S_3|$ (right axis) directly derived from Eq.\eqref{eq:S3} from a model without AC Stark shifts. They are compared to the red horizontal line which is the best estimate $|S_3|^{\mathrm{est}}$ (given in Fig.\ref{fig:ColorMaps_S3est}) derived from the model with AC Stark shifts. The red-shaded area corresponds to the 95\% confidence interval. The red error bars correspond to 1-$\sigma$ deviations considering the same level of uncertainty (see appendix \ref{DataAnalysis}).}
	\label{fig:E_and_S3_estimation}
\end{figure}  

For the quasi-circular, respectively elliptical, polarization the method leads to an estimated value for the third Stokes parameter of $|S_3|^{\mathrm{est}}=0.954\pm0.01$, respectively $|S_3|^{\mathrm{est}}=0.833\pm0.006$ (see appendix \ref{DataAnalysis} for more details on the uncertainty evaluation). For this estimation, we retain only the spectra in which the electric field amplitude is large enough for the 4 resonances to be clearly resolved: the last 4 data-sets for the quasi-circular case and the last 5 data-sets for the elliptical one. Notably for quasi-circular polarization, we note the importance of acquiring spectra with sufficient microwave power. Indeed, at low power the two central resonances are not resolved preventing us to estimate the actual, albeit small, deviation from circularity that is clearly visible for higher power (\textit{cf.} Fig.\ref{fig:ColorMaps_S3est}). Furthermore, our study highlights the need to account for off-resonant couplings to obtain an unbiased estimate of $|S_3|$ given that the non-zero linewidth of the resonances prevents us to resolve them at low power where Eq.\eqref{eq:S3} is valid. We illustrate this in Fig.\ref{fig:E_and_S3_estimation}: at high field amplitude, the direct estimation of $|S_3|$ from Eq.\eqref{eq:S3} is significantly biased (see appendix \ref{DataAnalysis} for more details on the uncertainties calculation). When the field decreases, this estimate tends towards the best estimate $|S_3|^{\mathrm{est}}$ from the aforedescribed method indicated by the horizontal red line in Fig. \ref{fig:E_and_S3_estimation}. In the elliptical case, the direct estimate appears to deteriorate for the lowest amplitudes. This stems from additional biases in the fitting procedure when the resonance profiles start to overlap each other and which could originate from the influence of the MOT dynamics \cite{Duverger2024Metrology} and the presence of the cooling light induced Autler-Townes doublet \cite{Duverger2024Metrology,bai2019single}. We also study the linearity of the field amplitude measured by the atoms as a function of the field applied at the input of the microwave horn antenna. We monitor the input power using a 50/50 splitter and a microwave average power sensor. For the same data-sets, the linear fits of Fig.\ref{fig:E_and_S3_estimation} show a linearity of the scale factor better than 1~\% (see also appendix \ref{DataAnalysis} and Fig.\ref{fig:E_relative_error}) from which we conclude that both parameters $|E|$ and $|S_3|$ are evaluated to this level.

For the quasi-linear polarization, we cannot directly apply the above method as we do not observe a lift of degeneracy for the range of applied power. Nevertheless, we notice a broadening of the resonances from typically 6~MHz to 10.4~MHz (1$\sigma$) while increasing the power. Those measured linewidths allow us to set an upper bound on the deviation to a linear polarization: $|S_3|^{\mathrm{est}} \leq 0.11$. In that case, the linearity of the scale factor still remains better than 1\% (see Fig. \ref{fig:E_relative_error}).

\section{Impact of the polarization and direction of the Rydberg beams}\label{rydberg_polarization}

The visibility of the trap-loss signal is directly proportional to the probability for the atoms to be excited in Rydberg states as the loss mechanism is based on their ionization by the surrounding black-body radiation \citep{Duverger2024Metrology}. We therefore derive the probability $P_{|g\rangle \rightarrow |m_{J''}\rangle}$ to excite an atom from the ground state $|g\rangle$ to a specific Rydberg state $|nS_{1/2},m_{J''}\rangle$ by summing on all possible excitation paths and considering a statistical mixture equally distributed among the ground state sub-levels $|g, m_F\rangle$ (see appendix \ref{Appendix:ProbabilityRydberg}).

\begin{figure}
    \includegraphics[width=0.95\linewidth]{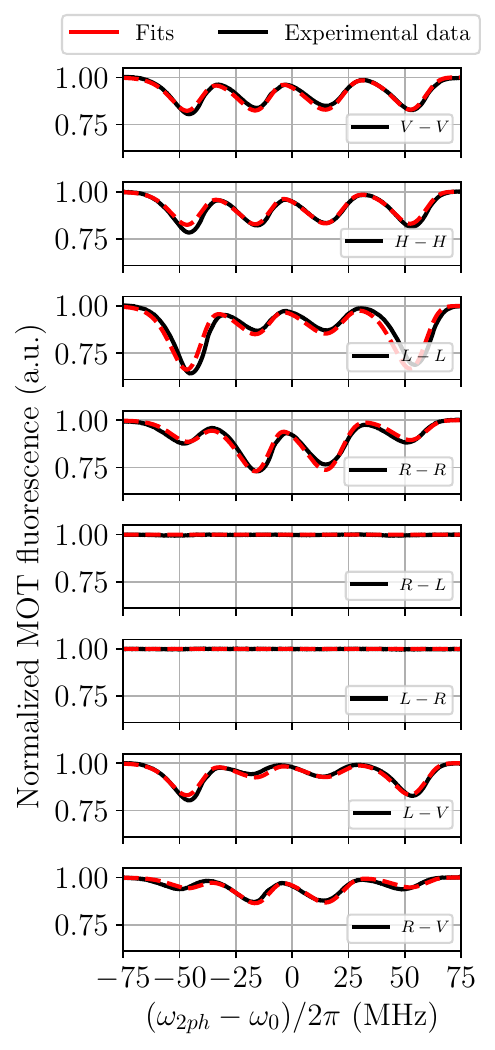}
    \caption{Trap-loss spectra for different Rydberg laser polarization configurations for $|E|=4.4$ V/m and $|S_3| = 0.833$. The states of polarization $\vec{\epsilon}^{\,(1)} -$ $\vec{\epsilon}^{\,(2)}$ (780~nm $-$ 480~nm) are denoted as V, H, L and R for vertically, horizontally, left-handed circularly and right-handed circularly polarized beam respectively. These states of polarization are given with respect to the direction of propagation of each beam. The experimental data are depicted in solid black lines while the dashed red lines correspond to the best adjustment by the theoretical model for parameters $\kappa=0.018,\rho= \pm 1.83~\mathrm{rad},\eta=\pm1.02~\mathrm{rad}$.}
\label{fig:polarrydberg}
\end{figure}

For the elliptical microwave polarization, \textit{i.e.} $|S_3|^{\mathrm{est}} = 0.833$, and for an applied field of $4.4$~V/m (corresponding to $20\log_{10}(E/E_{ref})=12.94$) at the atom position, we record several trap-loss spectra for 8 different polarization configurations $\{\vec{\epsilon}^{\,(1)} \, ; \,\vec{\epsilon}^{\,(2)} \}$ of the two Rydberg excitation beams. The spectra are plotted in Fig.\ref{fig:polarrydberg} and the polarizations are given with respect to the propagation axis of each beam in this figure. We clearly observe strong variations of the resonance amplitudes across the different configurations with identical values for eigenstates within the same subspaces as described in appendix \ref{Appendix:ProbabilityRydberg}, resulting in symmetric spectra with respect to the central frequency. Notably, the polarization of the Rydberg beams does not affect the resonance frequencies, and thus the estimation of $|S_3|$, as expected for a perturbative two-photon coupling. 

In order to compare the observed contrasts with the predictions from Eq.\eqref{Proba_2ph} of appendix \ref{Appendix:ProbabilityRydberg}, we first extract the resonance amplitudes by multi-peaks gaussian fits when the contrast is sufficient. Due to the surrounding environment of the atoms, the only unknown of the problem is the relative direction between the microwave direction of propagation at the atoms location, and the axis defined by the two counter-propagating Rydberg beams. Taking this into account, we minimize the difference between the observed and predicted amplitudes across all configurations with the propagation direction and a proportionality factor $\kappa$ between excitation probabilities and resonance amplitudes as free parameters (see appendix \ref{DataAnalysis} and Fig.\ref{fig:AnglesErrorMap} for more details). 

The predictions are depicted in dashed red lines in Fig.\ref{fig:polarrydberg} where the resonance frequencies and widths are set to their average values across the different configurations. We find a good agreement with the measurements for a pair of angles $\left\{ \rho = \pm 1.83~\mathrm{rad},\eta = \pm 1.02~\mathrm{rad} \right\}$ and an overall proportionality factor $\kappa=0.018$. The angles $\rho$ and $\eta$ correspond to the angles of the microwave wave vector at the atomic cloud position relative to the direction orthogonal to the Rydberg beams: $\rho$, respectively $\eta$, lies in, respectively out, the plane defined by the initial RF wave vector $\left\{ \rho_i = - 0.44 ~\mathrm{rad},\eta_i = 0~\mathrm{rad} \right\}$ and the Rydberg beams (see Fig. \ref{fig:Exp_setup}). The observed deflection of the RF wave vector at the position of the atoms is probably the result of the presence of several reflecting and diffracting elements in our experimental setup. The remaining ambiguities are related to the symmetries of the problem and to the unknown sign of $|S_3|$.

From the spectra shown in Fig.\ref{fig:polarrydberg}, two linear parallel polarization states for the Rydberg laser beams appear to be the most suited configuration as it gives rise to visible resonances of similar amplitudes with a weak dependence on the relative angle with the microwave wave vector. On the contrary, the L $-$ R and R $-$ L coupling configurations are the least suited as the visibilities of the resonances are essentially zero because of selection rules ($m_{J''} \neq m_J \pm 2$). The 4 other configurations preferentially enhance the coupling towards one of the two sub-spaces leading to different amplitudes for the two inner and the two outer resonances with an amplitude ratio that depends strongly on the relative direction of the microwave.

\section{Conclusion}

In this work, we have demonstrated and evaluated the capability of a Rydberg-based sensor to measure simultaneously the amplitude and the polarization ellipticity of arbitrary microwave fields based on SI-traceable frequency measurement using a simple $nS_{1/2} \leftrightarrow n'P_{1/2}$ transition. We have proposed a simple physical model to show how the energy's spectrum of the coupled \{atom - RF field\} system encodes information about the two quantities under study with a scale factor depending on fundamental and atomic constants only. We have established the reliability of this method over a power dynamic range of typically 20 dB and emphasized the need to take into account the shifts induced by off-resonant coupling with neighboring Rydberg states in order to retain an accurate estimation of $|S_3|$, with a residual error below 1\% over this range. The capability given by cold atoms to perform sub-wavelength measurements of the MW field extends the possibility to resolve slight deviations to purely circular or linear polarization states at high MW power. A natural and direct improvement of this method would be to narrow the linewidth of the different transitions by mitigating the dominant broadening mechanisms: for instance by performing the Rydberg excitation when the cooling light is turned off. This would significantly enhance the accuracy and the dynamic range over which the method is efficient.Then it would also be possible to resolve distinct Zeeman transitions from residual ambient static magnetic fields. Another natural extension of this method would be to record and analyze spectra for different $J \leftrightarrow J'$ transitions \cite{Kjaergaard2026}. For instance, by considering a $nS_{1/2} \leftrightarrow n'P_{3/2}$ transition, one obtains a spectrum with 2 degenerate and $|S_3|$-independent eigenenergies along with 4 $|S_3|$-dependent ones. The same amount of information regarding the state of polarization can thus be extracted in that case, with the advantage that the 4 eigenenergies are non-degenerate for a circular polarization. Finally, the simple approach presented in this work enables vector field characterization without requiring multiple local oscillators or complex heterodyne setups, offering a promising route for SI-traceable microwave sensing applications in radar, telecommunications or space science \cite{Allinson2026}. Moreover, the dependence of the amplitudes of the spectroscopic features with the direction of arrival could be harnessed to give some information on the latter for an incoming MW signal in a single-point measurement.

Finally, cold atom technologies also offer the possibility to realize arrays of localized atomic probes embedded in a compact sensor head. This could also be harnessed to acquire information on the angle of the incoming wave vector or to perform multi-parameter estimation in order to retrieve a complete characterization of any incoming RF field.

\begin{acknowledgments}
We acknowledge funding from a French government grant managed by the Agence Nationale de la Recherche in the France 2030 framework, with the reference ANR-23-PETQ-0004, and from the project ANR-22-CE47-0009-03. Romain Granier acknowledges a PhD fellowship from Ecole Normale Supérieure Paris-Saclay in the CDSN program. Anthony El Bekai acknowledges a PhD fellowship from the graduate school of Université Paris-Saclay.
\end{acknowledgments}

\newpage


\section{Appendix}

\subsection{Electric dipole matrix elements and polarization definition}\label{Appendix:A}

For a radio-frequency electric field that couples $|61S_{1/2}\rangle\equiv|s\rangle$ and $|61P_{1/2}\rangle\equiv |p\rangle$ (the other quantum numbers are omitted from the notation for simplicity), we consider the four magnetic sub-levels of these states $|s, m_j=\pm1/2\rangle$ and $|p, m_j=\pm1/2\rangle$, which can be written as:
\[
\begin{cases}
    |s,\frac{1}{2}\rangle \hspace{0.25cm} = |l=0,m_l=0,m_s=\frac{1}{2}\rangle \\
    |s,-\frac{1}{2}\rangle = |l=0,m_l=0,m_s=-\frac{1}{2}\rangle  \\
    |p,\frac{1}{2}\rangle \hspace{0.25cm} = \frac{|l=1, m_l=0, m_s=\frac{1}{2}\rangle-\sqrt{2}|l=1, m_l=1, m_s=-\frac{1}{2}\rangle}{\sqrt{3}} \\
    |p,-\frac{1}{2}\rangle = \frac{\sqrt{2}|l=1, m_l=-1, m_s=\frac{1}{2}\rangle-|l=1, m_l=0, m_s=-\frac{1}{2}\rangle}{\sqrt{3}} 
\end{cases}
\]
The electric-dipole Hamiltonian is $\hat{H}(t)=-\hat{\vec{d}} \cdot \vec{E}(t)= -\hat{\vec{d}} \cdot \vec{E} \cos(\omega t) = - \frac{1}{2} \hat{\vec{d}} \cdot \vec{E} \left(e^{-i\omega t} + e^{i\omega t}\right)$ and in such a frame, the relevant dipole matrix elements to take into account are $\langle s,\pm \frac{1}{2}|\hat{\vec{d}} \cdot \vec{E}|p,\pm\frac{1}{2}\rangle$ with $\hat{\vec{d}}=-e \hat{\vec{r}}$. Since the dipole operator does not act on spin, it only couples states that share the same spin projection $m_s$. Consequently, the problem reduces to the evaluation of $\langle l=0, m_l=0 | \hat{\vec{d}}\cdot\vec{E} | l=1, m_l=\left\{-1,0,+1\right\}\rangle$.

Defining the electric field as
\begin{equation}
        \vec{E} = 
        \begin{cases}
        E_x = E \cos \left(\frac{\theta}{2}\right) \\
        E_y = E \sin \left(\frac{\theta}{2}\right) e^{-i\phi} \quad ,\\
        E_z = 0
        \end{cases}
        \label{eq:fieldform}
\end{equation}
it can be rewritten in the spherical basis as $\vec{E} = E_0 \vec{\epsilon}_{0} + E_- \vec{\epsilon}_{+} + E_+ \vec{\epsilon}_{-}$ (with the usual definition $\vec{\epsilon}_{0} = \vec{e}_{z}$, $\vec{\epsilon}_{+} = -\frac{1}{\sqrt{2}}(\vec{e}_{x}+i\vec{e}_{y})$, $\vec{\epsilon}_{-} = \frac{1}{\sqrt{2}}(\vec{e}_{x}-i\vec{e}_{y})$), leading to $\vec{E} = E_0 \vec{\epsilon}_{0} + E_- \vec{\epsilon}_{+} + E_+ \vec{\epsilon}_{-}$; $E_0 =E_z$, $E_+ =\frac{E_x+iE_y}{\sqrt{2}}$ and $E_- =-\frac{E_x-iE_y}{\sqrt{2}}$. One also has $\vec{r} = r_0 \vec{\epsilon}_{0} + r_- \vec{\epsilon}_{+} + r_+ \vec{\epsilon}_{-}$, with $r_{0} = r \cos \theta$, $r_{-} = - \frac{r \sin \theta}{\sqrt{2}} e^{-i \varphi}$ and $r_{+} = \frac{r \sin \theta}{\sqrt{2}} e^{i \varphi}$. Denoting the radial part of the dipole matrix element $\mathcal{R} = \int_{r=0}^{+\infty} R^*_{61,0,\frac{1}{2}}R_{61,1,\frac{1}{2}}r^3 dr$, one can derive:
\begin{eqnarray}{}
    \langle s,\frac{1}{2} | \hat{d}_0 E_0 |p,\frac{1}{2}\rangle =& -\frac{1}{3}eE_0\mathcal{R}  &= - \hbar
    \Omega_0 \\
    \langle s,-\frac{1}{2} | \hat{d}_0 E_0 |p,-\frac{1}{2}\rangle =& \frac{1}{3}eE_0\mathcal{R} &= \hbar
    \Omega_0  \\
    \langle s,\frac{1}{2} | \hat{d}_+ E_+ |p,-\frac{1}{2}\rangle =& -\frac{\sqrt{2}}{3}eE_+\mathcal{R} &=
    -\hbar \Omega_+ \\
    \langle s,-\frac{1}{2} | \hat{d}_- E_- |p,\frac{1}{2}\rangle =& \frac{\sqrt{2}}{3}eE_-\mathcal{R} &=
    \hbar \Omega_- 
\end{eqnarray}
where we call $d = - \frac{e \mathcal{R}}{3}$ and define $\hbar \Omega_0 = -dE_0$, $\frac{\hbar \Omega_+}{\sqrt{2}} = -dE_+$, $\frac{\hbar \Omega_-}{\sqrt{2}} = -dE_-$. $\left(\Omega_0, \Omega_+, \Omega_-\right)$ are associated with $\left(E_0, E_+, E_-\right)$ for $\left(\pi, \sigma_+, \sigma_-\right)$ transitions respectively.  In the most general case, we find $4$ different possible eigenvalues sorted as $\lambda_{12} < \lambda_{11} < \lambda_{21} < \lambda_{22}$.

The polarization can be represented on the Poincaré sphere using the Stokes parameters $\left(S_0, S_1 , S_2, S_3\right)$ or via the geometric parameters of the polarization ellipse (azimuth angle $\alpha$, ellipticity $\varepsilon \in \left[-\frac{\pi}{4},\,\frac{\pi}{4}\right]$). These quantities are linked as follows $S_1=\cos \theta = \cos (2\varepsilon) \cos (2\alpha)$, $S_2=\sin \theta \cos \phi = \cos (2\varepsilon) \sin (2\alpha)$, $S_3=\sin \theta \sin \phi = \sin (2\varepsilon)$.

Within this framework, we can express the Rabi frequencies $\Omega_{\pm}$ as a function of $S_3$ and $|E|$ as
\begin{equation}
	\Omega_{\pm}=\pm\frac{d|E|\sqrt{1\pm S_3}}{\hbar}
	\label{eq:Omegapm}
\end{equation}
\subsection{Energy shifts from off-resonant couplings and refined Hamiltonian}\label{Appendix:B}

The preceding development did not consider the energy shifts that can be induced by neighboring states. As the principal quantum number $n$ increases, the atomic energy levels become increasingly closely spaced, resulting in a richly structured energetic landscape. In practice, a resonant radio-frequency field coupling two levels will also produce AC Stark shifts on the adjacent levels due to off-resonant couplings. Consequently, each involved level $|61S_{1/2},1/2\rangle$, $|61S_{1/2},-1/2\rangle$, $|61P_{1/2},1/2\rangle$, $|61P_{1/2},-1/2\rangle$ will undergo AC Stark shifts $\delta^{S,\frac{1}{2}}_{LS}$, $\delta^{S,-\frac{1}{2}}_{LS}$, $\delta^{P,\frac{1}{2}}_{LS}$ and $\delta^{P,-\frac{1}{2}}_{LS}$ due to off-resonant RF-coupling with neighboring states to which it can be coupled.

These terms are calculated by summing energy shifts of the form:
\begin{equation}
	\mathrm{sgn}\left(\delta_{LS}^{i \leftrightarrow j}\right)\frac{\Omega_{\pm,ij}^2}{4\Delta_{ij}}
\end{equation}
The couplings $\Omega_{\pm,ij}$ are evaluated for all transitions $i\leftrightarrow j$ allowed by selection rules, where the difference of modulus between $\Omega_{+,ij}$ and $\Omega_{-,ij}$ is the signature of the $|S_3|$-dependency (see Eq.\ref{eq:Omegapm}).
We define the detunings $\Delta_{ij}$ as positive quantities. We take the convention of a plus, respectively minus, sign for an increasing, respectively decreasing, energy shift accordingly to blue or red detuned off-resonant coupling.

We restrict ourselves to the states given in Table \ref{tab:lightshift}, which were found to contribute for more than 100 kHz for a reference Rabi frequency of 100 MHz.

\begin{table}[h]
	\centering
	\begin{tabular}{ccc}
		\toprule
		Transition &$\Delta_{ij}/2\pi$& Radial dipole \\
		&  (MHz) & ($ea_0$) \\
		\hline
		$61 S_{\frac12} \leftrightarrow 61 P_{\frac32}$ & $437$ &$3814$ \\
		$61 S_{\frac12} \leftrightarrow 60 P_{\frac12}$& $2054$&$3528$ \\
		$61 S_{\frac12} \leftrightarrow 60 P_{\frac32}$& $1593$&$3592$ \\
		$61 P_{\frac12} \leftrightarrow 62 S_{\frac12}$& $1149$&$3652$ \\
		$61 P_{\frac12} \leftrightarrow 60 D_{\frac32}$& $5881$&$4632$ \\
		$61 P_{\frac12} \leftrightarrow 59 D_{\frac32}$ & $7398$&$2658$ \\
		\hline
	\end{tabular}
	\caption{Relevant transitions with the radial part of the dipole matrix element and their detunings $\Delta_{ij}$ in units of MHz which is taken as positive by convention.}
	\label{tab:lightshift}
\end{table}

\begin{figure}
	\includegraphics[width=0.95\linewidth]{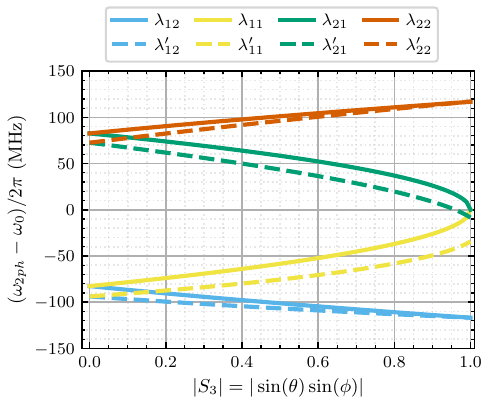}
	\caption{Comparison of eigenenergies of $\hat{H}$ (solid lines) and $\hat{H}'$ (dashed lines) in unit of MHz as a function of $|S_3|$ for $|E|=10$ V/m}
	\label{fig:eigenenergies_noLSvsLS}
\end{figure}

All these energy shift terms will appear on the diagonal, resulting in the following refined Hamiltonian $\hat{H}'$:

\[
\hat{H}' = \frac{\hbar}{2}
\renewcommand{\arraystretch}{1.6}
\setlength{\arraycolsep}{6pt}
\raisebox{2ex}{$
\begin{array}{c@{\hspace{-10pt}}c}
&
\begin{array}{cccc}
{|s,\frac12}\rangle &
{|p,-\frac12}\rangle &
{|s,-\frac12}\rangle &
{|p,\frac12}\rangle
\end{array}
\\[1ex]
\begin{array}{c}
\langle{s,\frac12}| \\
\langle{p,-\frac12}| \\
\langle{s,-\frac12}| \\
\langle{p,\frac12}|
\end{array}
&
\left[
\begin{array}{cccc}
2\delta^{S,\frac12}_{LS} & \Omega_+ & 0 & 0 \\
\Omega_+^* & 2\delta^{P,-\frac12}_{LS} & 0 & 0 \\
0 & 0 & 2\delta^{S,-\frac12}_{LS} & -\Omega_- \\
0 & 0 & -\Omega_-^* & 2\delta^{P,\frac12}_{LS}
\end{array}
\right]
\end{array}
$}
\]
resulting in the following expressions of the eigenenergies relative to the 2 independent 2-level systems:
\begin{equation}
	E^{+}_{\pm} = \frac{\delta^{S,\frac12}_{LS}+\delta^{P,-\frac12}_{LS}}{2}\pm\frac12\sqrt{|\Omega_+|^2+(\delta^{S,\frac12}_{LS}-\delta^{P,-\frac12}_{LS})^2}
\end{equation}

\begin{equation}
	E^{-}_{\pm} = \frac{\delta^{S,-\frac12}_{LS}+\delta^{P,\frac12}_{LS}}{2}\pm\frac12\sqrt{|\Omega_-|^2+(\delta^{S,-\frac12}_{LS}-\delta^{P,\frac12}_{LS})^2}
\end{equation}
We illustrate the impact of AC Stark shifts by plotting the eigenenergies of $\hat{H}$ and $\hat{H}'$ in Fig.\ref{fig:eigenenergies_noLSvsLS}.

For this first estimation of $|E|$ at the atom cloud position, we neglect the contribution of AC Stark shifts: indeed, for the highest power levels, they generate maximum differential shifts of 17.1 MHz ; $\{18.2;8.9\}$ MHz ; $\{36.5;19.7\}$ MHz for the linear ; circular ; elliptical cases, which, compared to the AT-splittings of 156.5 MHz ; $\{30;195.2\}$ MHz ; $\{82.1;272.1\}$ MHz, yield relative errors of $6\times10^{-3}$ ; $5\times10^{-3}$ ; $1\times10^{-2}$ on the estimation of $|E|$ (where the two values are associated to each of the sub-spaces of dimension 2).

\subsection{Optical two-photon excitation rate}\label{Appendix:C}

We want to express the transition strengths that couple the ground state to the Rydberg dressed states. For this, as discussed in the main text, we need to calculate the Rabi frequencies between the ground state magnetic sublevels $|g\rangle = |5S_{1/2}, F=2, m_F\rangle$ with the quantum number $m_F \in [-2,2]$ to the states $|r\rangle = |61S_{1/2}, m_{J''}=\pm \frac{1}{2}\rangle$ for all possible transitions $\pi$, $\sigma_+$, $\sigma_-$ for the Rydberg lasers labeled $q_1$ and $q_2$. This excitation is performed  via the intermediate states $|e\rangle = |5P_{3/2}, F', m_{F'}\rangle$, $F' \in \{1,2,3\}$ with a detuning $\Delta/2\pi$ from the $|5P_{3/2}\rangle$ state.

Under our experimental conditions the single-photon Rabi frequencies can be employed to evaluate the effective two-photon Rabi frequency of a three-level ladder system. Because the one-photon detuning satisfies $\Delta \gg \Omega_i, \Gamma_i$ with $\Omega_i$ the single-photon Rabi frequencies and $\Gamma_i$ the corresponding transition linewidths (including all broadening mechanisms) the intermediate state remains essentially unpopulated. In this regime the dynamics may be reduced to an effective two-level system by adiabatic elimination of the intermediate level \cite{shore2008coherent}. The resulting two-photon Rabi frequency for the transition $|g\rangle \rightarrow |r\rangle$ is obtained by summing over all possible excitation paths with intermediate states $|e\rangle$ for a given $m_F$ \cite{steck2007quantum}:
\begin{equation}
\Omega_{|g\rangle \rightarrow |r\rangle} = \sum_{|e\rangle} \frac{\Omega^1_{|g\rangle \rightarrow |e\rangle} \Omega^2_{|e\rangle \rightarrow |r\rangle}}{2\Delta_e}
\end{equation}
where the exponents 1 and 2 refer to electric fields $E_{1,2}$ and $\Delta_e$ accounts for the hyperfine structure  $E_\mathrm{hfs}(F'))$ of the intermediate state. 

We finally compute the two-photon Rabi coupling using the Wigner-Eckart Theorem as:
\begin{equation}
\begin{aligned}
\Omega_{m_F, m_{J''}, q_{1}, q_{2}} &= 
\sum_{F'}\sum_{F''} 
\frac{E_1 E_2}{2 \hbar^2 (\Delta - E_\mathrm{hfs}(F'))} \\[0.2cm]
& \times 
(-1)^{q_2 + m_F + J'' + J' + J + I + F + F''}  \\[0.2cm]
& \times \sqrt{(2 F + 1)(2 F' + 1)(2 F'' + 1)} \\[0.2cm]
& \times
\begin{pmatrix} J'' & I & F'' \\ m_{J''} & m_{F''}-m_{J''} & -m_{F''} \end{pmatrix} \\[0.2cm]
& \times
\begin{pmatrix} F' & 1 & F'' \\ -m_{F'} & -q_2 & m_{F''} \end{pmatrix}
\begin{pmatrix} F & 1 & F' \\ -m_F & -q_1 & m_{F'} \end{pmatrix} \\[0.2cm]
& \times 
\left\{
\begin{matrix} F' & 1 & F'' \\ J'' & I & J' \end{matrix}
\right\}
\left\{
\begin{matrix} F & 1 & F' \\ J' & I & J \end{matrix}
\right\} \\[0.2cm]
& \times
\langle 5, S_{1/2} || e a_0 r|| 5, P_{3/2} \rangle
\langle 5, P_{3/2} || e a_0 r || 61, S_{1/2} \rangle
\end{aligned}
\end{equation}

\subsection{Probability of excitation to Rydberg states}\label{Appendix:ProbabilityRydberg}

 In the absence of coupling, we describe the atoms as a statistical mixture with equal probabilities of all 5 magnetic sub-levels $m_F$ of the hyperfine ground state $|g \rangle = |5S_{1/2}, F=2\rangle$. For each ground state sub-level $|g, m_F \rangle$, we treat the effective two-photon laser coupling between $|g, m_F \rangle$ and the dressed Rydberg states, namely the eigenstates of $\hat{H}$ described in section \ref{theory}, as a small perturbation to $\hat{H}$. In this situation, the frequencies at which the coupling is the most efficient are given by the eigenfrequencies of $\hat{H}$, and the corresponding coupling strength is given by the coupling strength between $|g, m_F \rangle$ and the corresponding eigenstate of $\hat{H}$. Because the coupling is a two-photon transition with the intermediate state $|i \rangle = |5P_{3/2}\rangle$, the transition to the Rydberg state $|n'P_{1/2}\rangle$ is forbidden by selection rules such that the coupling between $|g, m_F \rangle$ and a given eigenstate of $\hat{H}$ will only depend on the projection of the latter on the $\big\{|nS_{1/2},m_{J''}=\pm\frac{1}{2}\rangle \big\}$ manifold. Finally, we note that the eigenstates of $\hat{H}$ are either equal superpositions of $|nS_{1/2}, m_{J''}=+\frac{1}{2}\rangle$ and $|n'P_{1/2}, m_{J''}=-\frac{1}{2}\rangle$ for a first manifold, or equal superpositions of $|nS_{1/2}, m_{J''}=-\frac{1}{2}\rangle$ and $|n'P_{1/2}, m_{J''}=+\frac{1}{2}\rangle$ for a second manifold. If we now write $E_1$ and $E_2$ the amplitudes of the coupling lasers and $\vec{\epsilon}^{\,(1)}$ and $\vec{\epsilon}^{\,(2)}$ their unitary polarization vectors, Rabi frequencies of the form $\Omega_{m_F,m_{J''},\alpha,\beta}$ can be calculated using $E_1$ and $E_2$ (see appendix \ref{Appendix:C}) for all possible values of the ground state magnetic sublevels $m_F \in \{-2,-1,0,+1,+2\}$, the $|nS_{1/2}\rangle$ magnetic sublevels $m_{J''} \in \{\pm1/2\}$ and the type of transition induced by each laser, labeled by $\alpha$ and $\beta$ with possible values $\pi$, $\sigma_+$ or $\sigma_-$. To obtain the probability of excitation between the ground state and any eigenstate of $\hat{H}$, we finally write the polarization vectors of the two coupling lasers $\vec{\epsilon}^{\,(1)}$ and $\vec{\epsilon}^{\,(2)}$ in the spherical basis whose quantization axis is aligned with the microwave direction of propagation as $\vec{\epsilon}^{\,(1)}=\sum_\alpha c_\alpha^{(1)} \vec{\epsilon}_\alpha$ and $\vec{\epsilon}^{\,(2)}=\sum_\beta c_\beta^{(2)} \vec{\epsilon}_\beta$, and sum over all possible excitation paths and all probabilities for all five ground state sublevels, leading to:

\begin{equation}\label{Proba_2ph}
	P_{|g\rangle \rightarrow |m_{J''}\rangle} \propto \sum_{m_F} \left| \sum_{\alpha,\beta} c_{\alpha}^{(1)}c_{\beta}^{(2)} \Omega_{m_F,m_{J''},\alpha,\beta}  \right|^2
\end{equation}

\subsection{Data Analysis}\label{DataAnalysis}

To take into account potential profile distortions due to dynamical effect during the frequency sweep, we use multi-peak asymmetric Gaussian fits \cite{Duverger2024Metrology}.

For evaluating the confidence interval on the estimation of $|S_3|$ we proceed as follow: we estimate the uncertainties within a 95\% confidence interval by considering an overall standard deviation of 1.36 MHz for the quasi-circular case, respectively 0.75 MHz for the elliptical case, on each resonance frequency across all powers. These 2 values correspond to the 1-$\sigma$ dispersion of the differences between the experimental data and the model. These differences take into the systematic errors from the experiment and the fitting procedure as the statistical error obtained from the fits are found to be negligible ($<0.25$ MHz peak-value). We then use the covariance matrix obtained from the \textit{curve\_fit} function of \textit{scipy.optimize} python library to estimate the 95\% confidence interval. Note that we also observe a constant systematic shift of $\leqslant 1.5$ MHz between the spectrum with microwave OFF or ON. This shift is due to different dynamical distortion of the resonance profiles of different contrasts. We neglect this shift since it has no influence on the estimation of $|S_3|$, being the same for all resonances frequencies.

\begin{figure}
	\includegraphics[width=0.95\linewidth]{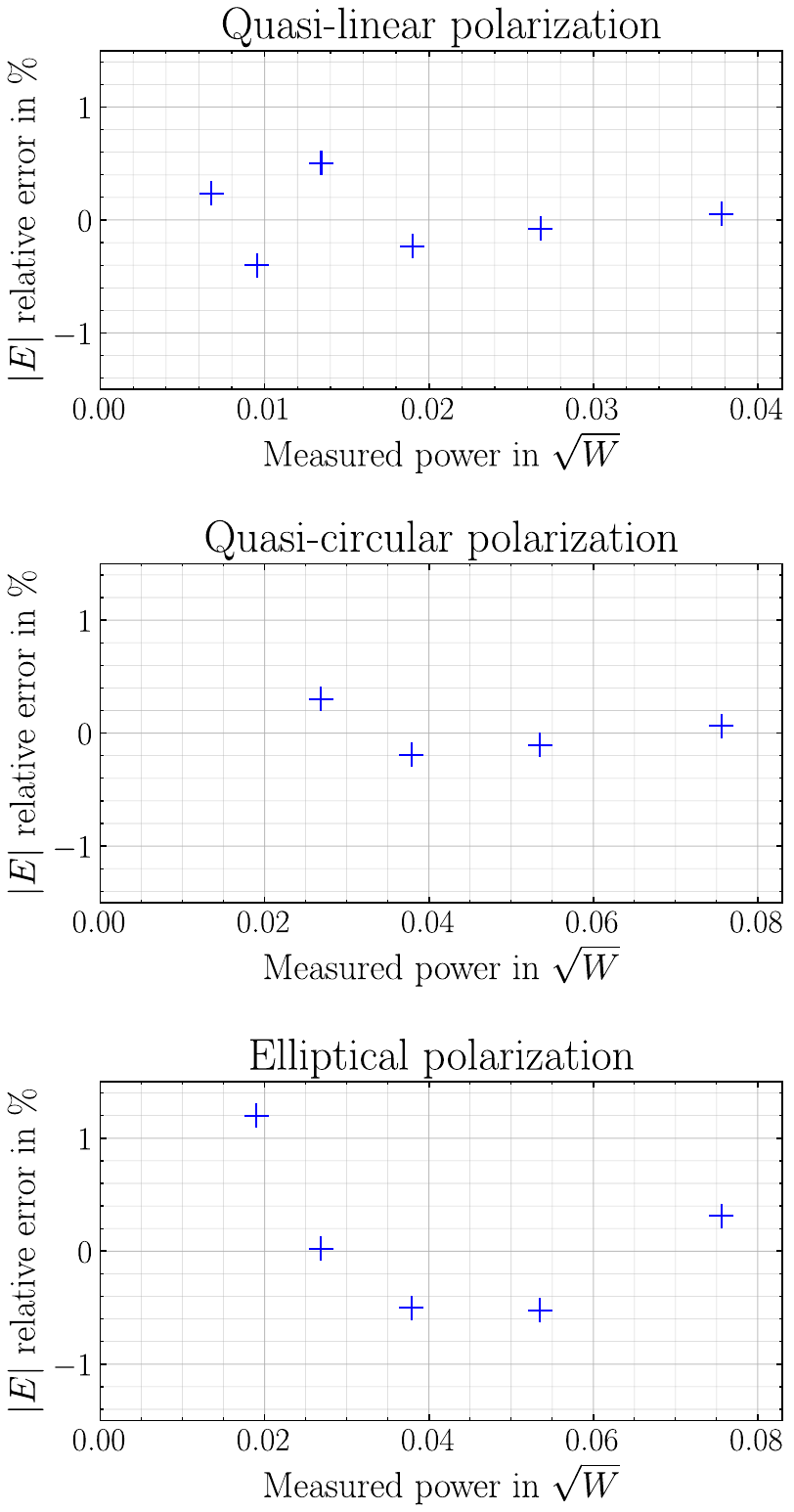}
	\caption{Residues of linear fits of electric field amplitudes $|E|$ as a function of the measured microwave power. The linear adjustments of the 3 data-sets give: $248.7\pm 0.5 ~ V.m^{-1}.W^{-1/2} + 0.08\pm 0.01 ~ V.m^{-1}$ for quasi-linear polarization, $112.0\pm 0.39 ~ V.m^{-1}.W^{-1/2} + 0.09\pm 0.02 ~ V.m^{-1}$ for quasi-circular polarization and $159.2\pm 1 ~ V.m^{-1}.W^{-1/2} + 0.16\pm 0.05 ~ V.m^{-1}$ for elliptical polarization.}
	\label{fig:E_relative_error}
\end{figure}

For the red errorbars on $|S_3|$ of Fig.\ref{fig:E_and_S3_estimation}, we consider the same deviation of 1.36 MHz for the quasi-circular case, respectively 0.75 MHz for the elliptical case, on all resonance frequencies. We then propagate this error using Eq.\eqref{eq:S3} at each RF power. These levels of uncertainties lead to errorbars on the blue points corresponding to different field amplitudes that are much smaller than the marker sizes. The linear fit results and the associated residues are given in Fig.\ref{fig:E_relative_error}.

\begin{figure}
	\includegraphics[width=0.95\linewidth]{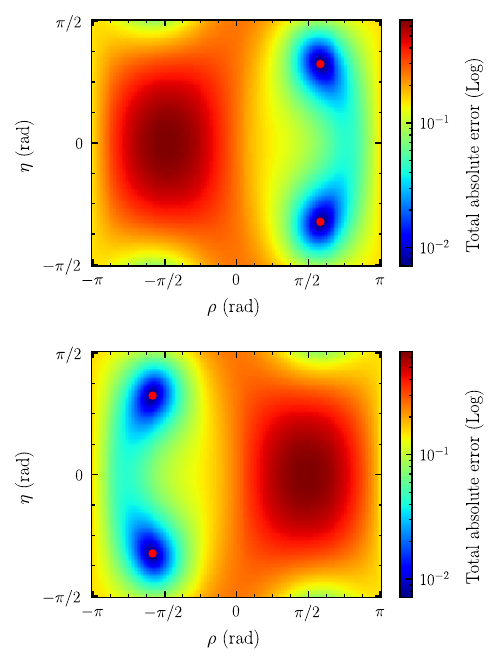}
	\caption{Error map as a function of the microwave direction of propagation at the atom cloud position relative to the Rydberg laser beam axis ($\rho,\eta$) resulting from the evaluation of the resonance amplitudes of Fig.\ref{fig:polarrydberg} for $\kappa=0.018$. Here, the two figures correspond to the two possible definitions of $m_{J''}$. The red points indicate the minima.}
	\label{fig:AnglesErrorMap}
\end{figure}

For the visibilities as a function of the Rydberg beams polarization (\textit{cf.} Fig.\ref{fig:polarrydberg}) we minimize the quantity :
\begin{equation}
	\label{ErrorAmplitude}
	\sum_{k,l} \left( A_{k,l}^{\mathrm{exp}} - \kappa P_{|g\rangle \rightarrow |m_{J''}=\pm1/2\rangle} (\rho,\eta) \right)^2
\end{equation}
where $A_{k,l}^{\mathrm{exp}}$ are the resonance amplitudes normalized by the mean fluorescence level of each spectrum and $\kappa$ is a proportionality factor left as a free parameter. The indexes $k, l$ span over the different experimental configurations and the 4 resonances: note that the two projections $m_{J''}=\pm1/2$ correspond to either the two inner or two outer resonances. We plot in Fig.\ref{fig:AnglesErrorMap} the error map corresponding to the quantity defined in Eq.\eqref{ErrorAmplitude}.

\bibliography{biblio}

\end{document}